\documentclass[prl,twocolumn,superscriptaddress,showpacs,floatfix]{revtex4}
\usepackage{amsmath}
\usepackage{amssymb}
\usepackage{graphicx}
\usepackage[english]{babel}
\usepackage{latexsym}
\usepackage{graphics}
\usepackage{subfigure}

\def\be{\begin{equation}}
\def\ee{\end{equation}}
\def\bea{\begin{eqnarray}}
\def\eea{\end{eqnarray}}
\def\bse{\begin{subequations}}
\def\ese{\end{subequations}}

\def\be{\begin{eqnarray}}
\def\ee{\end{eqnarray}}

\begin{document}

\title{Tunable Spin-orbit Coupling and Quantum Phase Transition in a Trapped
Bose-Einstein Condensate}
\author{Yongping Zhang}
\affiliation{Department of Physics and Astronomy, Washington State University, Pullman,
WA, 99164 USA}
\author{Gang Chen}
\affiliation{Department of Physics and Astronomy, Washington State University, Pullman,
WA, 99164 USA}
\affiliation{State Key Laboratory of Quantum Optics and Quantum Optics Devices, College
of Physics and Electronic Engineering, Shanxi University, Taiyuan 030006, P.
R. China}
\author{Chuanwei Zhang}
\thanks{Corresponding author, email: cwzhang@wsu.edu}
\affiliation{Department of Physics and Astronomy, Washington State University, Pullman,
WA, 99164 USA}

\begin{abstract}
Spin-orbit coupling (SOC), the intrinsic interaction between a particle spin
and its motion, is responsible for various important phenomena, ranging from
atomic fine structure to topological condensed matter physics. The recent
experimental breakthrough on the realization of SOC for ultra-cold atoms
provides a completely new platform for exploring spin-orbit coupled
superfluid physics. However, the SOC strength in the experiment, determined
by the applied laser wavelengths, is not tunable. In this Letter, we propose
a scheme for tuning the SOC strength through a fast and coherent modulation
of the laser intensities. We show that the many-body interaction between
atoms, together with the tunable SOC, can drive a \textit{quantum phase
transition} (QPT) from spin-balanced to spin-polarized ground states in a
harmonic trapped Bose-Einstein condensate (BEC). This transition realizes
the long-sought QPT in the quantum Dicke model, and may have important
applications in quantum optics and quantum information. We characterize the
QPT using the periods of collective oscillations (center of mass motion and
scissors mode) of the BEC, which show pronounced peaks and damping around
the quantum critical point.
\end{abstract}

\maketitle



SOC plays a major role in many important condensed matter phenomena and
applications, including spin and anomalous Hall effects \cite{Xiao},
topological insulators \cite{Hasan}, spintronics \cite{Dassarma}, spin
quantum computation, \textit{etc}. In the past several decades, there has
been tremendous efforts for developing new materials with strong SOC and new
methods for tuning SOC with high accuracy for spin-based device applications
\cite{TuneSOC1,TuneSOC2}. However, the SOC strength in typical solid state
materials (e.g., $\sim 10^{4}$ m/s in semiconductors) is generally much
smaller than the Fermi velocity of electrons ($\sim 10^{6}$ m/s), and its
tunability is limited and inaccurate.

On the other hand, the recent experimental breakthrough on the realization
of SOC for ultra-cold atoms \cite{Lin} provides a completely new platform
for exploring SOC physics in both BEC \cite{Wu,Wang,Ho,Yongping,Hu,Santos}
and degenerate Fermi gases \cite{Zhang,DFG1,DFG2,DFG3}. In a degenerate
Fermi gas, such SOC strength can be at the same order as (or even larger
than) the Fermi velocity of atoms. Because of the strong SOC, spins are not
conserved during their motion and new exotic superfluids may emerge. For
instance, new ground state phases (e.g., stripes, phase separation, etc.)
may be observed in spin-orbit coupled BEC \cite{Wang,Ho,Yongping,Hu,Santos}\
and new topological excitations (e.g., Weyl \cite{DFG1} and Majorana \cite%
{Zhang} fermions) may appear in spin-orbit coupled Fermi gases. The
observation and application of these exciting phenomena require tunable SOC
for cold atoms. Unfortunately, the strength of the SOC in the experiment
\cite{Lin} and other theoretical proposals \cite%
{Ruseckas,Zhang2,Dalibard,Campbell} is not tunable because the SOC strength
is determined by the directions and wavelengths, not the intensities, of the
applied lasers.

In this Letter, we propose a scheme for generating tunable SOC for cold
atoms through a fast and coherent modulation of the Raman laser intensities
\cite{Grifoni}, which can be easily implemented in experiments. Such tunable
SOC for cold atoms provides a powerful tool for exploring new exotic Bose
and Fermi superfluid phenomena. Here we focus on a \textit{quantum phase
transition} (QPT) \cite{Sachdev} in a harmonic trapped BEC induced by the
many-body interaction between atoms and the tunable SOC strength. With the
increasing SOC strength, there is a sharp transition for the ground state of
the BEC from a spin balanced (\textit{i.e.}, equally mixed) phase to a spin
fully polarized phase beyond a critical SOC strength (\textit{i.e.}, the
quantum critical point). By mapping the spin-orbit coupled interacting BEC
to the well-known quantum Dicke model \cite{Dicke,Emary}, we obtain analytic
expressions for the quantum critical point and the corresponding scaling
behaviors for the QPT, which agree well with the numerical results obtained
from the mean-field Gross-Pitaevskii (G-P) equation for the BEC.

The realization of QPT in the Dicke model using the spin-orbit coupled BEC
opens the door for many significant applications in quantum optics, quantum
information, and nuclear physics \cite{DM1,DM2,DM3}. Previously the Dicke
model has been studied in several experimental systems \cite%
{Schoelkopf,Garraway}, especially atoms confined in an optical cavity.
However the coupling between atoms and optical cavity fields is very weak,
and the experimental observation of the QPT in the Dicke model only occurred
recently using the momentum eigenstates for a BEC confined in a cavity \cite%
{Baumann}. Compared with the cavity scheme, the spin-orbit coupled BEC
utilizes the many-body interaction between atoms and has the advantage of
essentially no dissipation, fully tunable parameters, very strong coupling,
and the use of atom internal states, thus provides an excellent platform for
exploring Dicke model related applications.

Finally, the QPT is characterized using collective oscillations of the BEC,
such as the center of mass (COM) motion and the scissors mode, where the
oscillation periods show pronounced peaks at the quantum critical point.
Furthermore, the oscillations of the BEC have regular periodic patterns in
both spin balanced and polarized phases, but show strong damping in the
transition region.

\begin{figure}[t]
\includegraphics[width=0.7\linewidth]{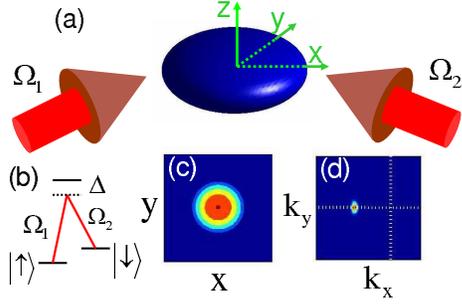}
\caption{(Color Online) An illustration of the experimental scheme for
realizing SOC for cold atoms \protect\cite{Lin}. (a) Laser setup. (b) Atom
laser coupling. (c) A typical density distribution of one spin component of
the BEC in the ground state. (d) Corresponding momentum distribution. The
vertical and horizontal dotted lines are $k_{x}=0$ and $k_{y}=0$
respectively.}
\label{setup}
\end{figure}

\textit{System and Hamiltonian}: The harmonic trapped BEC in consideration
is similar as that in the recent benchmark experiment \cite{Lin}. For
simplicity, we consider a two-dimensional (2D) BEC in the $xy$ plane with a
strong confinement (with a trapping frequency $\omega _{z}$) along the $z$
direction. Such 2D setup does not affect the essential physics because the $%
z $ direction is not coupled with the SOC. Two hyperfine ground states $%
\left\vert \uparrow \right\rangle \equiv |F=1,m_{F}=-1\rangle $ and $%
\left\vert \downarrow \right\rangle \equiv |F=1,m_{F}=0\rangle $ of $^{87}$%
Rb atoms define the spins of atoms, which are coupled by two Raman lasers
(with Rabi frequencies $\Omega _{1}$ and $\Omega _{2}$) incident at a $\pi
/4 $ angle from the $x$ axis, as illustrated in Fig. 1a and 1b.

The dynamics of the BEC are governed by the nonlinear G-P equation
\begin{equation}
i\hbar \partial \Phi /\partial t=\left( p^{2}/2m+V\left( \mathbf{r}\right)
+H_{S}+H_{I}\right) \Phi ,  \label{TGP}
\end{equation}%
under the dressed state basis $\left\vert \bar{\uparrow}\right\rangle =\exp
\left( i\mathbf{k}_{1}\cdot \mathbf{r}\right) \left\vert \uparrow
\right\rangle $, $\left\vert \bar{\downarrow}\right\rangle =\exp \left( i%
\mathbf{k}_{2}\cdot \mathbf{r}\right) \left\vert \downarrow \right\rangle $,
where $\mathbf{k}_{1}$ and $\mathbf{k}_{2}$ are the wavevectors of the
lasers. $\Phi =(\Phi _{\uparrow },\Phi _{\downarrow })^{T}$ is the
wavefunction on the dressed state basis and satisfies the normalization
condition $\int dxdy(|\Phi _{\uparrow }|^{2}+|\Phi _{\downarrow }|^{2})=1$.
The harmonic trapping potential $V\left( \mathbf{r}\right) =\frac{1}{2}%
m\omega _{y}^{2}(\eta ^{2}x^{2}+y^{2})$, where $\omega _{y}$ is the trapping
frequency in the $y$ direction, and $\eta =\omega _{x}/\omega _{y}$ is the
ratio of the trapping frequencies. $H_{S}=\gamma p_{x}\sigma _{z}+\hbar
\Omega \sigma _{x}/2$ is the coupling term induced by the two Raman lasers
with $\sigma _{z}$ and $\sigma _{x}$ as the\ Pauli matrices. The SOC
strength $\gamma =\hbar k_{L}/m$, $k_{L}=\left\vert \mathbf{k}_{1}-\mathbf{k}%
_{2}\right\vert /2=\sqrt{2}\pi /\lambda $, and $\lambda $ is the wavelength
of the Raman lasers. The Raman coupling constant $\Omega =\Omega _{1}\Omega
_{2}^{\ast }/2\Delta $ with $\Delta $ as the detuning from the excited
state. The mean field nonlinear interaction term $H_{I}=diag\left(
g_{\uparrow \uparrow }|\Phi _{\uparrow }|^{2}+g_{\uparrow \downarrow }|\Phi
_{\downarrow }|^{2},g_{\uparrow \downarrow }|\Phi _{\uparrow
}|^{2}+g_{\downarrow \downarrow }|\Phi _{\downarrow }|^{2}\right) $, where
the inter- and intra-spin interaction constants $g_{\uparrow \uparrow
}=g_{\uparrow \downarrow }=4\pi \hbar ^{2}N(c_{0}+c_{2})/ma_{z}$ and $%
g_{\downarrow \downarrow }=4\pi \hbar ^{2}Nc_{0}/ma_{z}$, $c_{0}$ and $c_{2}$
describe the corresponding \textit{s}-wave scattering lengths \cite{Ho2}, $N$
is the atom number, and $a_{z}=\sqrt{2\pi \hbar /m\omega _{z}}$.

Because the SOC strength $\gamma $ is determined by the laser wavevector $%
k_{L}$, the SOC energy can be comparable to or even larger than other energy
scales (e.g., the Raman coupling $\Omega $) in the BEC. In a Fermi gas, $%
\gamma $ can be larger than the Fermi velocity of atoms. Unfortunately, due
to the same reason, $\gamma $ cannot be easily adjusted in experiments,
which significantly restricts the applications of the SOC in cold atoms.

\textit{Tunable SOC for cold atoms}-- We propose a scheme for tuning the SOC
strength $\gamma $ through a fast and coherent modulation of the Raman
coupling $\Omega =\Omega _{0}+\widetilde{\Omega }\cos (\omega t)$ that can
be easily realized in experiments by varying the Raman laser intensities
\cite{Note}. Here the modulation frequency $\omega $ is chosen to be much
larger than other energy scales in Eq. (\ref{TGP}). In this case, the
Hamiltonian in Eq. (\ref{TGP}) can be transformed to a time-independent one
using a unitary transformation $\psi =\exp [i\widetilde{\Omega }\sin (\omega
t)\sigma _{x}/(2\omega )]\Phi $. After a straightforward calculation with
the elimination of the fast time-varying part in the Hamiltonian \cite%
{Eckardt,Lignier}, the nonlinear G-P equation (\ref{TGP}) becomes
\begin{equation}
i\hbar \partial \psi /\partial t=\left[ p^{2}/2m+V\left( \mathbf{r}\right) +%
\bar{H}_{s}+\bar{H}_{I}\right] \psi ,  \label{TCPS}
\end{equation}%
where the Raman coupling becomes $\bar{H}_{s}=\gamma _{eff}p_{x}\sigma
_{z}+\hbar \Omega _{0}\sigma _{x}/2$ with the effective SOC strength
\begin{equation}
\gamma _{eff}=\gamma J_{0}(\widetilde{\Omega }/\omega ).  \label{SOCS}
\end{equation}%
Here $J_{0}$ is the zero order Bessel function. Clearly, $\gamma _{eff}$ can
be tuned from the maximum $\gamma $ without the modulation to zero with a
strong modulation. The mean field interaction term $\bar{H}_{I}=\alpha
(|\psi _{\uparrow }|^{2}+|\psi _{\downarrow }|^{2})\psi +\beta \Gamma \psi $%
, $\alpha =g_{\uparrow \uparrow }$, $\beta =\left( g_{\downarrow \downarrow
}-g_{\uparrow \uparrow }\right) /2$ and $\Gamma $ is a $2\times 2$ matrix
whose elements are given by $\Gamma _{11}=|\psi _{\uparrow }|^{2}[\frac{3}{4}%
-J_{0}(\frac{\widetilde{\Omega }}{\omega })+\frac{1}{4}J_{0}(\frac{2%
\widetilde{\Omega }}{\omega })]+|\psi _{\downarrow }|^{2}[\frac{1}{4}-\frac{1%
}{4}J_{0}(\frac{2\widetilde{\Omega }}{\omega })]$, $\Gamma _{22}=|\psi
_{\uparrow }|^{2}[\frac{1}{4}-\frac{1}{4}J_{0}(\frac{2\widetilde{\Omega }}{%
\omega })]+|\psi _{\downarrow }|^{2}[\frac{3}{4}+J_{0}(\frac{\widetilde{%
\Omega }}{\omega })+\frac{1}{4}J_{0}(\frac{2\widetilde{\Omega }}{\omega })]$%
, and $\Gamma _{12}=\Gamma _{21}^{\ast }=-\frac{1}{4}[1-J_{0}(\frac{2%
\widetilde{\Omega }}{\omega })][\psi _{\uparrow }^{\ast }\psi _{\downarrow
}-\psi _{\uparrow }\psi _{\downarrow }^{\ast }]$.

We choose the physical parameters to be similar as those in the experiment
\cite{Lin}: $(\omega _{y},\omega _{z})=2\pi \times (40,400)$ Hz, $\eta =1$, $%
\lambda =804.1$ nm, $c_{0}=100.86a_{B}$, $c_{2}=-0.46a_{B}$ \cite{Widera}
with the Bohr radius $a_{B}$, $N=1\times 10^{4}$, $\omega =2\pi \times 4.5$
kHz. For the numerical simulation, we need a dimensionless G-P equation that
is obtained by choosing the units of the energy, length and time as $\hbar
\omega _{y}$, $\sqrt{\hbar /(m\omega _{y})}=1.7$ $\mu $m, and $1/\omega
_{y}=4$ ms, respectively. The dimensionless parameters in the G-P equation
become $\gamma =\sqrt{\hbar /(m\omega _{y})}k_{L}=9.37$, $\alpha =2N\sqrt{%
2\pi m\omega _{z}/\hbar }(c_{0}+c_{2})=495$ and $\beta =-N\sqrt{2\pi m\omega
_{z}/\hbar }c_{2}=1.14$.

\textit{Quantum phase transition:} The tunable SOC, in combination with the
many-body interaction between atoms, can drive a quantum phase transition
between different quantum ground states in a harmonic trapped BEC. Here the
ground state of the BEC is obtained numerically through an imaginary time
evolution of the G-P equation (\ref{TCPS}). A typical density profile of the
ground state is shown in Fig. 1c, which has a Thomas-Fermi shape, similar as
that in a regular BEC. However, the momentum distribution of the BEC has a
peak around the single particle potential minimum located at $\left(
K_{x},K_{y}\right) =\left( -K_{\min },0\right) $ (see Fig. 1d), where $%
K_{\min }=\sqrt{\gamma _{eff}^{2}-\Omega _{0}^{2}/4\gamma _{eff}^{2}}$ and
the degeneracy between $\pm K_{\min }$ is spontaneously broken.

To characterize the ground state of the spin-orbit coupled BEC, we calculate
the spin polarization $|\langle \sigma _{z}\rangle |=\left\vert \int d%
\mathbf{r}\left( \left\vert \psi _{\uparrow }\right\vert ^{2}-\left\vert
\psi _{\downarrow }\right\vert ^{2}\right) \right\vert $, and $\langle
\sigma _{x}\rangle =2$Re$\int d\mathbf{r}\psi _{\uparrow }^{\ast }\psi
_{\downarrow }$. Here we choose the absolute value of $\langle \sigma
_{z}\rangle $ because the two degenerate ground states at $\pm K_{\min }$
have opposite $\left\langle \sigma _{z}\right\rangle $ due to the
spin-momentum locking term $p_{x}\sigma _{z}$ and they are spontaneously
chosen in experiments. In Fig. 2a, we plot $|\langle \sigma _{z}\rangle |$
and $\langle \sigma _{x}\rangle $ with respect to $\gamma _{eff}$. For a
small $\gamma _{eff}$, the spin up and down atoms have an equal population,
thus $\left\langle \sigma _{z}\right\rangle =0$, $\langle \sigma _{x}\rangle
=-1$, \textit{i.e.}, the spin balanced phase. Beyond a critical point $%
\gamma _{eff}^{c}$, the spin population imbalance increases dramatically and
reaches the spin polarized phase $\left\langle \sigma _{z}\right\rangle =1$ (%
$\left\langle \sigma _{x}\right\rangle =0$) within a small range of $\gamma
_{eff}$. The spin balanced and spin polarized phases at small and large $%
\gamma _{eff}$ can be understood from the single particle Hamiltonian $\bar{H%
}_{s}$, where the Raman coupling $\Omega _{0}\sigma _{x}/2$ and SOC $\gamma
_{eff}p_{x}\sigma _{z}$ dominate at the small and large $\gamma _{eff}$
respectively. More numerical results show that the critical transition point
occurs at $\gamma _{eff}^{c}=\sqrt{\Omega _{0}/2}$.

\begin{figure}[tbp]
\includegraphics[width=0.7\linewidth]{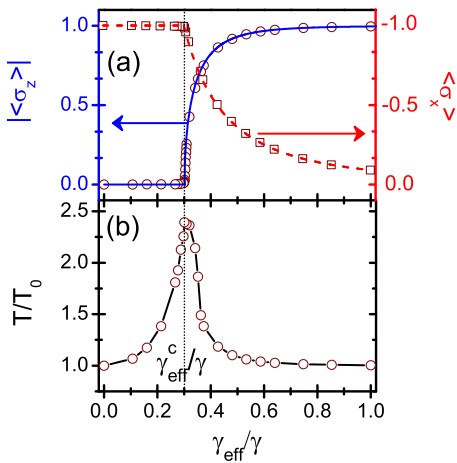}
\caption{(Color Online) Quantum phase transition with tunable $\protect%
\gamma _{eff}$. $\Omega _{0}=16$, $\protect\gamma =9.37$ is the bare SOC
strength without modulation. (a) Plot of the spin polarization $\left\vert
\left\langle \protect\sigma _{z}\right\rangle \right\vert $ and $%
\left\langle \protect\sigma _{x}\right\rangle $ in the ground state. The
blue and red lines are from the prediction of the Dicke Hamiltonian. The
circles and squares are from the numerical simulation of the G-P equation (%
\protect\ref{TCPS}). (b) Plot of the COM motion period $T$. The shift of the
harmonic trap $D=1$, $T_{0}=2\protect\pi /\protect\omega _{y}$.}
\label{QPT}
\end{figure}

The QPT can be understood by mapping the spin-orbit coupled BEC to a quantum
Dicke model. For an interacting BEC in a harmonic trap with a large atom
number $N$ (so that the mean field theory works), all atoms are forced to
occupy the same many-body ground state. Therefore the energy variation for
the change of the spin (e.g., spin flip) of one atom need be determined by
the coupling between the atom spin and the many-body ground state mode. This
is very different from an non-interacting BEC where atoms do not affect each
other, but the same as that for many atoms interacting with a single photon
mode in a cavity \cite{Dicke}. Treating the interacting many-body ground
state as a single mode composed of different harmonic trap modes, we can
map the Hamiltonian for the spin-orbit coupled BEC to%
\begin{equation}
H=\hbar \omega _{x}Na^{\dagger }a+\hbar \Omega _{0}S_{x}+\frac{\gamma _{eff}%
\sqrt{m\hbar \omega _{x}}}{\sqrt{2}}(a^{\dagger }-a)(S_{+}-S_{-}),
\label{DH}
\end{equation}%
which is similar to the Dicke model for two-level atoms coupled with a
cavity field \cite{Dicke}. $S_{x,y,z}$ are the large spins for all atoms, $%
S_{+}=S_{y}+iS_{z}$, $S_{-}=S_{y}-iS_{z}$, $a^{\dagger }a$ is a harmonic
trap mode, $a=\sqrt{m\omega _{x}/2\hbar }(x+ip_{x}/m\omega _{x})$. The
critical point for the QPT can be derived from the standard mean-field
approximation \cite{Emary}, yielding the relation $\gamma _{eff}^{c}=\sqrt{%
\Omega _{0}/2}$, which is exactly the same as that from numerically
simulating the G-P equation (\ref{TCPS}). Just beyond the critical point $%
\gamma _{eff}^{c}$, the Dicke model predicts that the scaling of the order
parameters is $|\langle \sigma _{z}\rangle |=2|S_{z}|/N=\sqrt{1-(\gamma
_{eff}^{c}/\gamma _{eff})^{4}}$, $\langle \sigma _{x}\rangle
=2S_{x}/N=-(\gamma _{eff}^{c}/\gamma _{eff})^{2}$ for $\gamma _{eff}\geq
\gamma _{eff}^{c}$, and $|\langle \sigma _{z}\rangle |=0$, $\langle \sigma
_{x}\rangle =-1$ for $\gamma _{eff}<\gamma _{eff}^{c}$. Such scaling
behaviors are confirmed in our numerical simulation of the G-P equation (see
Fig. 2a). The perfect match between numerical results from the G-P equation
and the predictions of the Dicke Hamiltonian shows the validity of the
mapping to the Dicke model.

We emphasize that the many-body interaction between atoms plays a critical
role in the QPT by forcing all atoms in a single mode. Without interaction,
numerical simulation of the G-P equation shows $\left\langle \sigma
_{z}\right\rangle =0$ in certain region of $\gamma _{eff}>\gamma _{eff}^{c}$%
, which disagrees with the prediction of the Dicke model. This disagreement
indicates that atoms in a non-interacting BEC do not response to the change
of $\gamma _{eff}$ collectively, although non-interacting and interacting
BECs share the same transition for the energy spectrum at $\gamma _{eff}^{c}$%
, which changes from one single minimum at $K_{x}=0$ to two minima at $\pm
K_{\min }$. While for interacting BECs with large atom numbers $N=4\times
10^{4}$ and $10^{6}$, we obtain exactly the same results as that in Fig. 2a,
which further confirm the validity of our mapping to the Dicke model in the
large $N$ limit.

\emph{Collective dynamics in BEC: the signature of QPT}: It is well-known
that various physical quantities may change dramatically around the quantum
critical point (\textit{i.e.}, critical phenomena), which provides
additional experimental signatures of the QPT. We focus on two types of
collective dynamics of the ground state of the BEC: the COM motion and the
scissors mode induced by a sudden shift or rotation of the harmonic trapping
potential, respectively. In a regular BEC without SOC, the COM motion is a
standard method to calibrate the harmonic trapping frequency because the
oscillation period depends only on the trapping frequency \cite{Stringari}
and is not affected by other parameters such as nonlinearity, shift
direction and distance, etc.

\begin{figure}[tbp]
\includegraphics[width=1\linewidth]{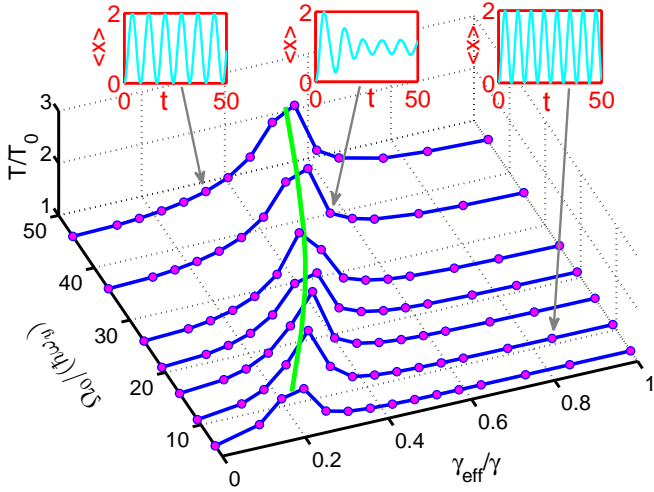}
\caption{(Color Online) Plot of the COM motion period $T$ versus $\protect%
\gamma _{eff}$ and $\Omega _{0}$. The insets show the corresponding $%
\left\langle x\left( t\right) \right\rangle $. The green line is the
theoretical prediction $\protect\gamma _{eff}^{c}=\protect\sqrt{\Omega _{0}/2%
}$ from the Dicke model. The circles are the numerical results from the G-P
equation (\protect\ref{TCPS}). $T_{0}=2\protect\pi /\protect\omega _{y}$, $%
D=1$. }
\label{COMT}
\end{figure}

\begin{figure}[b]
\includegraphics[width=1\linewidth]{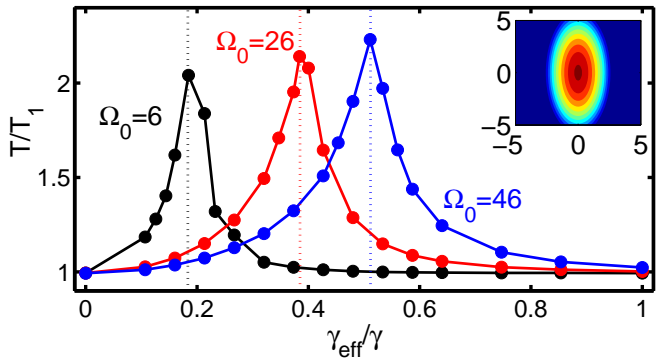}
\caption{Plot of the scissors mode oscillation period $T$ with respect to $%
\protect\gamma _{eff}$ for three different $\Omega _{0}$. $T_{1}=2\protect%
\pi /\protect\omega _{y}\protect\sqrt{\protect\eta ^{2}+1}$ is the
oscillation period without SOC, $\protect\eta =\protect\sqrt{5}$, $\protect%
\theta =4^{\circ }$. The inset is a typical density distribution of the
ground state of the BEC. The dotted lines show the corresponding quantum
critical points predicted by the Dick model. }
\label{scissormode}
\end{figure}
We numerically integrate the G-P equation (\ref{TCPS}) and calculate the COM
$\langle \mathbf{r}(t)\rangle =\int dxdy(|\psi _{\uparrow }(t)|^{2}+|\psi
_{\downarrow }(t)|^{2})\mathbf{r}\left( t\right) $. The COM motion strongly
depends on the direction of the shift $\vec{D}$ of the harmonic trap. When $%
\vec{D}$ is along the \textit{y} direction, the period of the COM motion
along the $y$ direction is $T_{0}=2\pi /\omega _{y}$ and not affected by $%
\gamma _{eff}$, while the COM motion in the \textit{x} direction disappears (%
\textit{i.e.}, $\langle x\rangle =0$). Here the COM period $T$ is obtained
through the Fourier analysis of $\langle \mathbf{r}\left( t\right) \rangle $%
. The physics is very different when $\vec{D}$ is along the \textit{x}
direction, where $\langle y\left( t\right) \rangle =0$ as expected, but $%
\langle x\left( t\right) \rangle $ depends strongly on $\gamma _{eff}$, as
shown in Fig. 2b. In Fig. 3, we also plot $T$ as a function of $\gamma
_{eff} $ and $\Omega _{0}$. Without SOC ($\gamma _{eff}=0$), $T=T_{0}$, the
period for a regular BEC, as expected. $T$ increases with $\gamma _{eff}$ in
the spin balanced phase, but decreases when spin starts to be polarized,
leading to a sharp peak at the quantum critical point $\gamma _{eff}^{c}$.
The oscillation of $\langle x\left( t\right) \rangle $ in the spin balanced
phase is completely dissipationless, while a strong damping occurs in a
small range of $\gamma _{eff}$ beyond $\gamma _{eff}^{c}$ (see the inset in
Fig. 3). Far beyond $\gamma _{eff}^{c}$, the oscillation becomes regular
again with the period $T=T_{0}$ because the ground state has only one
component in this region. The peak and the damping of the oscillation around
$\gamma _{eff}^{c}$ provide clear experimental signatures for the QPT.
Moreover, $T$ also depends on the magnitude $D$ of the shift near the
critical point $\gamma _{eff}^{c}$: the larger $D$, the smaller $T$.

Another collective dynamics, the scissors mode \cite{Odelin}, shows a
similar feature as the COM motion. The scissors mode can be excited by a
sudden rotation of the asymmetric trapping potential (i.e., $\eta \neq 1$)
by an angle $\theta $, which induces an oscillation of the quantity $\langle
xy\rangle =\int dxdy(|\psi _{\uparrow }(t)|^{2}+|\psi _{\downarrow
}(t)|^{2})xy$. Without SOC, the period of the scissors mode is $T_{1}=$ $%
2\pi /\sqrt{\omega _{x}^{2}+\omega _{y}^{2}}$ \cite{Odelin}, as observed in
experiments \cite{Marago}. In Fig. 4, We plot the oscillation period $T$
with respect to $\gamma _{eff}$ for three different $\Omega _{0}$. We have
confirmed that the same QPT occurs for $\left\vert \left\langle \sigma
_{z}\right\rangle \right\vert $ and $\left\langle \sigma _{x}\right\rangle $
of the ground state in this asymmetric potential with the quantum critical
point $\gamma _{eff}^{c}=\sqrt{\Omega _{0}/2}$, as predicted by the Dicke
model. Similar as the COM motion, we observe the peak and damping of the
oscillation around $\gamma _{eff}^{c}$. Far beyond $\gamma _{eff}^{c}$, the
oscillation period is $T_{1}$. Similar as the dependence of the COM motion
on the shift distance $D$, the angle $\theta $ also influences the period of
the scissors mode near $\gamma _{eff}^{c}$: the smaller $\theta $, the
larger $T$.

In summary, we show that the SOC strength in the recent breakthrough
experiment for realizing SOC for cold atoms can be tuned through a fast and
coherent modulation of the applied laser intensities. Such tunable SOC
provides a powerful tool for exploring spin-orbit coupled superfluid physics
in future experiments. By varying the SOC strength, the many-body
interaction between atoms can drive a QPT from spin balanced to spin
polarized ground states in a harmonic trapped BEC, which realizes the
long-sought QPT in the Dicke model and may have important applications in
quantum information and quantum optics.

We thank helpful discussion with Peter Engels, Li Mao and Chunlei Qu. This
work is supported by DARPA-YFA (N66001-10-1-4025), ARO (W911NF-09-1-0248),
and NSF-PHY (1104546). Gang Chen is also supported by the 973 program under
Grant No. 2012CB921603 and the NNSFC under Grant No. 11074154.


\begin{thebibliography}{99}
\bibitem{Note} For $\widetilde{\Omega }>\Omega _{0}$, $\Omega $ changes sign
at certain time, which can be achieved by applying a $\pi $ phase shift on
one Raman laser.
\end{thebibliography}
\end{document}